\documentclass[12pt]{article}
\usepackage{graphicx}
\begin{document}

\pagestyle{empty}

\noindent
{\bf Amplitude Variations in Pulsating Red Giants. II. Some Systematics}

\bigskip

\noindent {\bf John R. Percy}

\noindent Department of Astronomy and Astrophysics, and Dunlap Institute
of Astronomy and Astrophysics, University of Toronto, Toronto ON Canada
M5S 3H4, john.percy@utoronto.ca

\smallskip

\noindent {\bf Jennifer Laing}

\noindent Department of Astronomy and Astrophysics, University of Toronto,
Toronto ON Canada M5S 3H4, jen.laing@mail.utoronto.ca

\bigskip

{\bf Abstract}  In order to extend our previous studies of the unexplained
phenomenon of cyclic amplitude variations in pulsating red giants,
we have used the AAVSO time-series analysis package VSTAR to analyze
long-term AAVSO visual observations of 50 such stars, mostly Mira stars.
The relative amount of the variation, typically a factor of 1.5, and the
time scale of the variation, typically 20-35 pulsation periods, are not
significantly different in longer-period, shorter-period, and carbon stars
in our sample, and they also occur in stars whose period is changing
secularly, perhaps due to a thermal pulse.  The time scale of the 
variations is similar to that in
smaller-amplitude SR variables, but the {\it relative} amount of the variation
appears to be larger in smaller-amplitude stars, and is therefore more
conspicuous.  The cause of the amplitude variations remains unknown.

\medskip

\noindent
AAVSO keywords = AAVSO International Database; photometry, visual; pulsating variables; giants, red; period analysis; amplitude analysis

\medskip

\noindent
ADS keywords = stars; stars: late-type; techniques: photometric; methods: statistical; stars: variable; stars: oscillations

\medskip

\noindent
{\bf 1. Introduction}

\smallskip

Percy and Abachi (2013) showed that, in almost all pulsating red giants
(PRGs), the pulsation {\it amplitude} varied by a factor of up to 10,
on a time scale of 20-40 pulsation periods.  The authors were initially
concerned that the variation might be an artifact of wavelet analysis,
but it can be confirmed by Fourier analysis of individual sections
of the dataset.  Similar amplitude variations were found in pulsating
red supergiants (Percy and Khatu 2014) and yellow supergiants (Percy
and Kim 2014).  There were already sporadic reports in the literature
of amplitude variations in PRGs (e.g. Templeton {\it et al.} 2008, Price and
Klingenberg 2005), but these stars tended to be the rare few which also
showed large changes in period, and which may be undergoing thermal pulses
(Uttenthaler {\it et al.} 2011).  Furthermore: it is well known that stars such
as Mira do not repeat exactly from cycle to cycle.  Percy and Abachi
(2013), however, was the first {\it systematic} study of this phenomenon.  Since these
amplitude variations remain unexplained, we have examined the behavior
of more PRGs, to investigate some of the systematics of this phenomenon.

We have analyzed samples of {\it large-amplitude} PRGs, mostly Mira stars,
in each of four groups: A: 17 shorter-period stars; B: 20 longer-period
stars; C: 15 carbon stars; D: 8 stars with significant secular period
changes (Templeton {\it et al.} 2005).  The stars in groups A, B, and C were
drawn randomly from among the 547 studied by Templeton {\it et al.} (2005) and
which did not show significant secular period changes.  As did Templeton
{\it et al.} (2005), we used visual observations from the American Association
of Variable Star Observers (AAVSO) International Database.  We did not
analyze stars for which the data were sparse, or had significant gaps.
Note that Templeton {\it et al.} (2005) specifically studied Mira variables,
which, by definition, have full ranges greater than 2.5 in visual light --
an arbitrary limit.

\medskip

The purposes of this paper are: (1) to present our analyses of these
50 PRGs, and (2) to remind the astronomical community, once again,
that the amplitude variations in PRGs require an explanation.

\medskip

\noindent
{\bf 2. Data and Analysis}

\smallskip

We analyzed visual observations from the AAVSO International Database
(AID: Kafka 2017) using the AAVSO's VSTAR software package (Benn 2013).
It includes both a Fourier and wavelet analysis routine; we used primarily
the latter.  For each star, we noted the Modified Julian Date MJD(1)
after which the data were suitable for analysis
 -- not sparse, no significant gaps.  From the WWZ wavelet plots, we
determined the maximum (Amx), minimum (Amn), and average (\={A})
amplitude, the number of cycles N of amplitude increase and decrease,
and the average length L of these cycles.  See Percy and Abachi (2013)
for a discussion of these quantities and their uncertainties; N and
therefore L can be quite uncertain because the cycles are irregular,
and few in number, especially if they are long.  This is doubly true for the
few stars in which the length of the dataset is shorter than average.
The maximum and minimum amplitudes are also uncertain since they are
determined over a limited interval of time.

We then calculated the ratio of L to the pulsation period P, the ratio
of maximum to minimum amplitude, the difference $\Delta$A between the
maximum and minimum amplitude, and the ratio of this to the average
amplitude \={A}.  The periods were taken from the VSX catalog, and rounded
off; the periods of stars like these ``wander" by several percent, due
to random cycle-to-cycle fluctuations.  All this information is listed
in Tables 1-4.  In the ``Notes" column, the symbols are as follows:
``s" -- the data were sparse in places; ``g" -- there were one or more gaps in
the data (but not enough to interfere with the analysis); ``d" -- the
star is discordant in one or more graphs mentioned below, but there
were no reasons to doubt the data or analysis; asterisk (*) -- see Note
in Section 3.2.  Note that the amplitudes that we determine and list
are ``half-amplitudes" rather than the full ranges i.e. they are the
coefficient of the sine function which fit to the data.


\medskip

\noindent
{\bf 3. Results}

\smallskip

We plotted L/P, Amx/Amn, and $\Delta$A/\={A} against period for each of
the four groups of stars A,B,C, and D.  There was no substantial trend in any case,
except as noted below (Figures 1-3).
We therefore determined the mean M and standard error of the mean
SEM, for each of the three quantities, for each of the four groups.
These are given in Table 5.  We also flagged any outliers in the graphs,
and reexamined the data and analysis.  If there was anything requiring
comment, that comment is given in Section 3.2.

In stars which are undergoing large, secular period changes, possibly as a
result of a thermal pulse, the size and length of the amplitude variation
cycles is marginally larger, but this may be partly due to the difficulty
of separating the cyclic and secular variations.  Note that cyclic
variations in amplitude are present during the secular ones in these stars.

We also found that, for the shorter-period stars, \={A} increased
with increasing period (Figure 4), but this is a well-known correlation.  The
very shortest-period PRGs have amplitudes of only hundredths of a magnitude.
There was no trend in amplitude for the longer-period stars.

The {\it relative} amount of variation in amplitude is slightly larger in
shorter-period, smaller-amplitude stars (Figure 5).  This is consistent
with the results of Percy and Abachi (2013), as discussed in Section 4.

The \={A} for the carbon stars are systematically lower than for
the oxygen stars (Figures 2 and 3).  Again, this is well-known; in the oxygen stars,
the visual amplitude is amplified by the temperature sensitivity of
TiO bands, which are not present in carbon stars.  Note also that the
carbon stars have longer periods, since they are in a larger, cooler,
and more highly evolved state.

\begin{figure}
\begin{center}
\includegraphics[height=7cm]{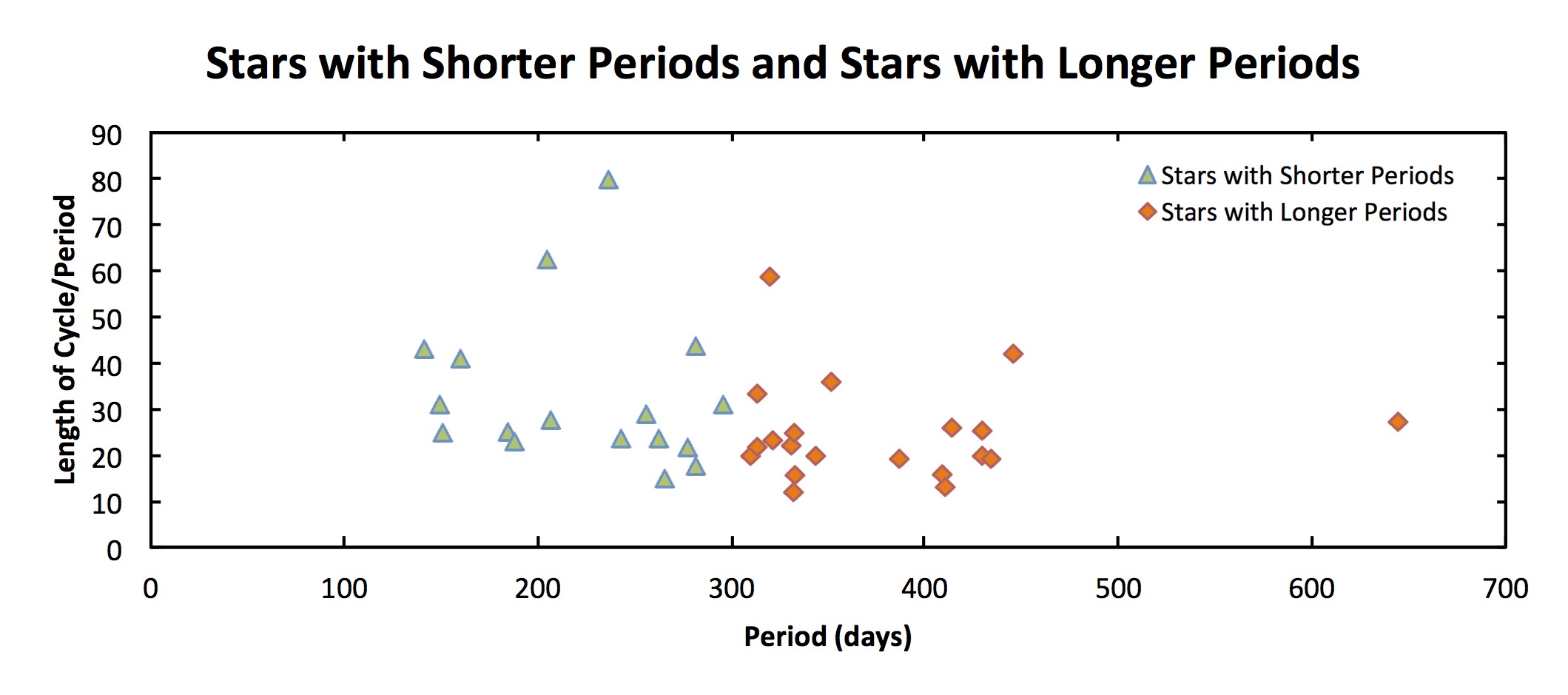}
\end{center}
\caption{The lengths of the cycles of amplitude increase and decrease, in
units of the pulsation period, as a function of pulsation period.  At most, 
there is a slight downward trend, which may be partly due to the fact that
the cycles may be more difficult to detect in shorter-period, smaller-amplitude
stars.}
\end{figure}

\begin{figure}
\begin{center}
\includegraphics[height=7cm]{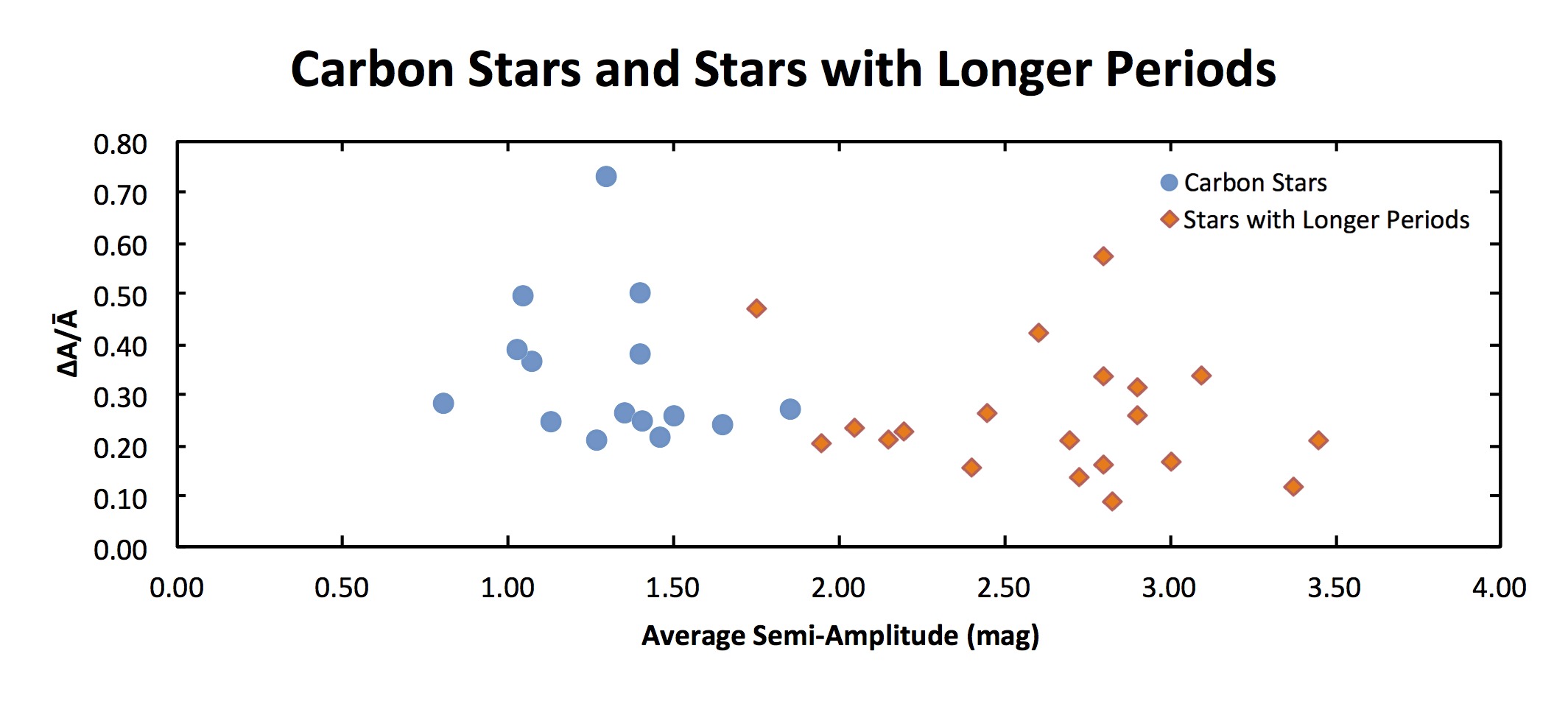}
\end{center}
\caption{The variation in visual amplitude, relative to the average visual
amplitude, as a function of average visual amplitude, for carbon stars
(blue filled circles) and non-carbon stars (red filled diamonds). 
The difference is not significant to the 3$\sigma$level (Table 5).}
\end{figure}

\begin{figure}
\begin{center}
\includegraphics[height=7cm]{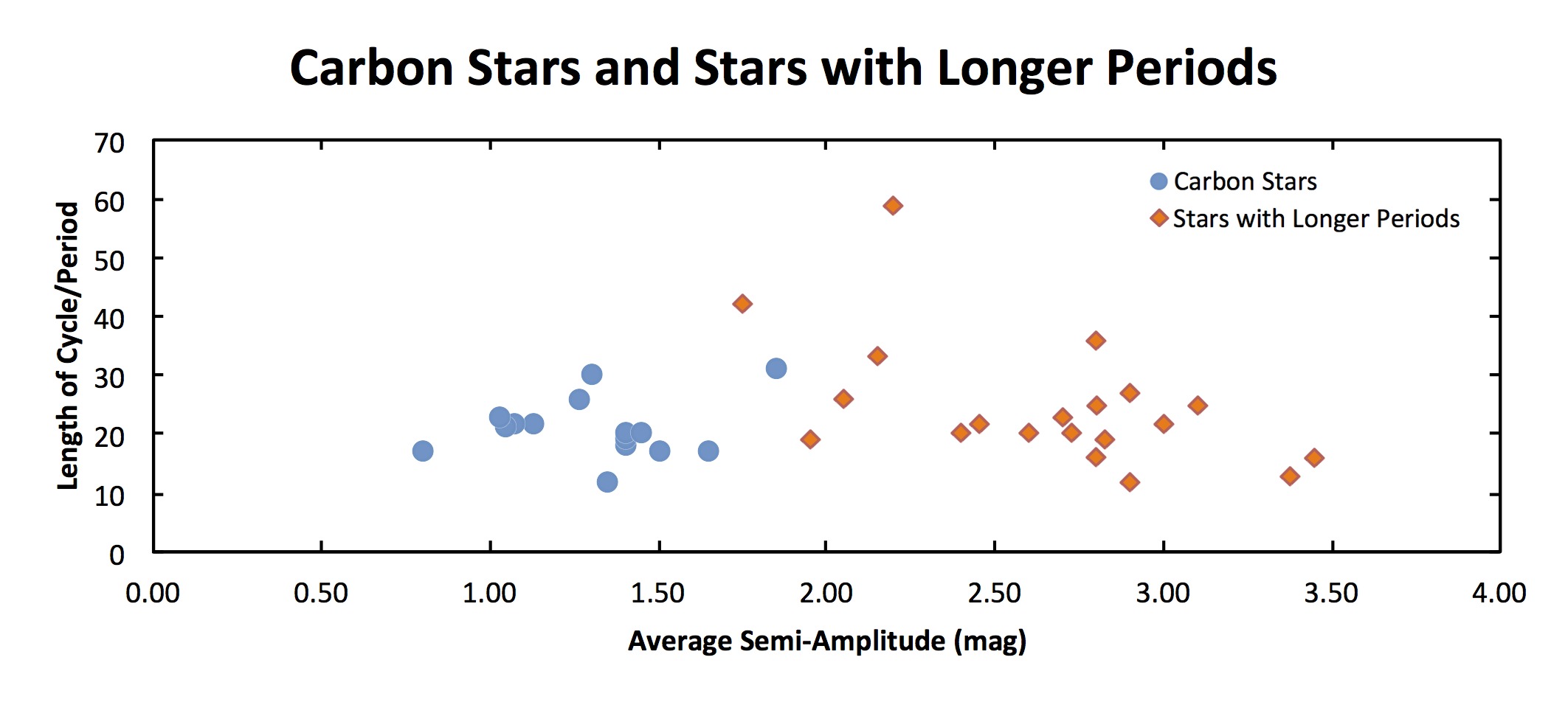}
\end{center}
\caption {The lengths of the cycles of amplitude increase and decrease, in
units of the pulsation period, as a function of average visual amplitude,
for carbon stars (blue filled circles) and non-carbon stars (red filled
diamonds). There is no trend.  The visual amplitudes of the carbon
stars are systematically smaller, as is well-known.}
\end{figure}
\medskip

\begin{figure}
\begin{center}
\includegraphics[height=7cm]{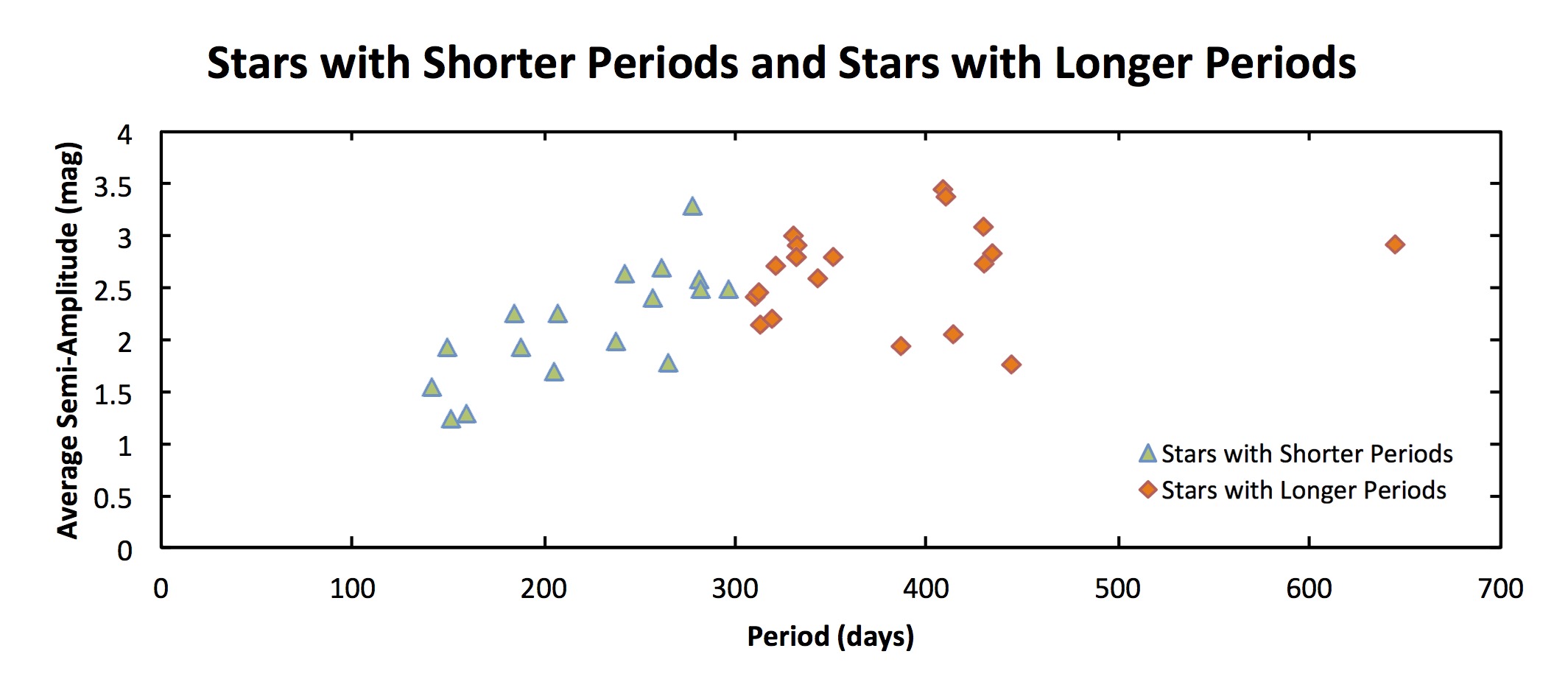}
\end{center}
\caption{The average visual pulsation amplitude as a function of pulsation
period, for shorter-period and longer-period Miras. The amplitude increases
with period, up to about 300 days (this is a continuation of a well-known
trend), and then levels off.}
\end{figure}

\begin{figure}
\begin{center}
\includegraphics[height=7cm]{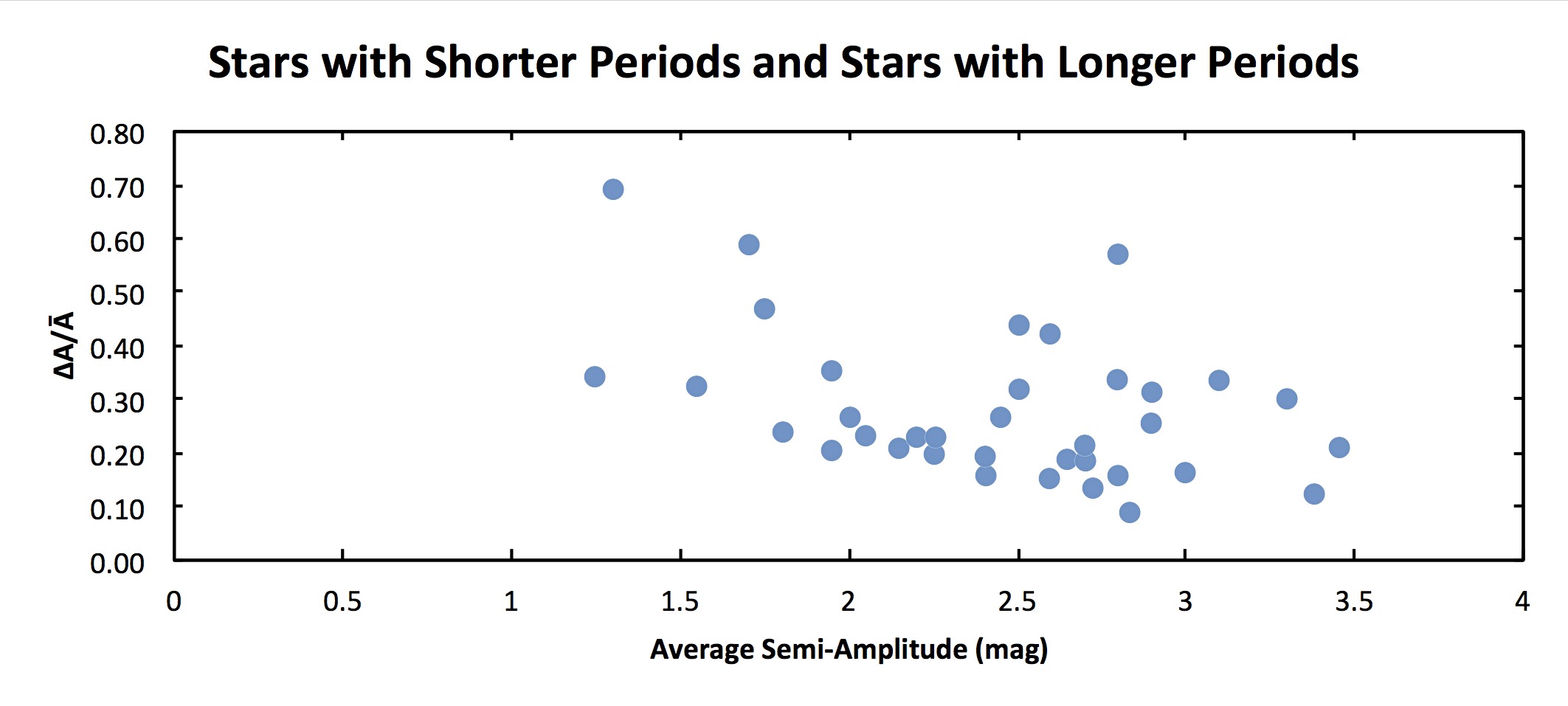}
\end{center}
\caption{The variation in visual amplitude, relative to the average visual amplitude,
as a function of average visual amplitude.  There is a downward trend.
This trend is consistent with the results of Percy and Abachi (2013), who
found values of typically 0.5 to 2.0 for stars with average amplitudes of 
1.0 down to 0.1.}
\end{figure}

\begin{figure}
\begin{center}
\includegraphics[height=7cm]{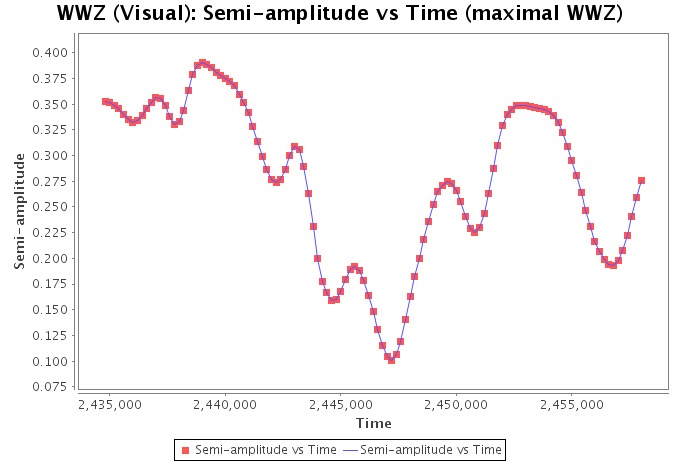}
\end{center}
\caption{Semi-amplitude versus time for RY Cam, a smaller-amplitude PRG.
Percy and Abachi (2013) estimated N conservatively at 1.5 but, based on
our subsequent experience, we would estimate N = 6.7.  This figure
therefore illustrates both the significant amplitude variation, and the
somewhat subjective estimate of N and therefore L.}
\end{figure}

\medskip

\noindent
{\bf 3.1 Stars with Secular Amplitude Variations}

\smallskip

Although our main interest was in the cyclic variations in pulsation
amplitude, the secular variations in amplitude are also of interest, though they
have already been studied and discussed by other authors, as mentioned
in the Introduction.  We performed a quick wavelet analysis of the 547
Miras in Templeton {\it et al.}'s (2005) paper, to identify stars in which
{\it secular} amplitude variations might dominate the cyclic ones.  Of the 21
stars whose period varied secularly at the three-sigma level or greater,
four (T UMi, LX Cyg, R Cen, and RU Sco) seemed to show such secular amplitude variations.
There were no other stars in Templeton {\it et al.}'s (2005) sample which
showed {\it strong} secular variations.  Note that, in each case, cyclic
amplitude variations were superimposed on the secular ones.

\medskip

\noindent
{\bf 3.2 Notes on Individual Stars}

\smallskip

This section includes notes on two kinds of stars: the ones for which the
data or analysis required comments, and ones which appear to be outliers
in some of the graphs that we have plotted.

\smallskip

{\it R Cen}: this star has a secular decrease in amplitude, and period, so
it is not surprising that the star is discordant in some of the relationships;
see also Templeton {\it et al.} (2005).

{\it T Dra:} this star has unusually large cyclic variations in amplitude.

{\it R Lep}: this star has unusual large variations in mean magnitude.

{\it RZ Sco:} this star, with a relatively short period, has a secular
change in period, but only at the 3$\sigma$ level (Templeton {\it et al.} 2005).

{\it Z Tau}: this star is exceptional in that it is an S-type star.  Also: its
light curve shows non-sinusoidal variations, and flat minima suggestive
that the variable may have a faint companion star.  Indeed, SIMBAD lists
two faint stars within 5 arc seconds of Z Tau.  This star is discussed by
Templeton {\it et al.} (2005).

\medskip

\noindent
{\bf 4. Discussion}

\smallskip

Percy and Abachi (2013) obtained a median value of L/P = 44 for 28
monoperiodic smaller-amplitude PRGs.  They calculated the median,
in part because there were a few stars with very large values of L/P.
We have reanalyzed those stars, and realized that Percy and Abachi (2013)
adopted a more conservative definition for amplitude variations.
Figure 6 shows an example of this: for the smaller-amplitude PRG RY Cam,
Percy and Abachi (2013) estimated N = 1.5 whereas, based on our subsequent
experience, we would estimate N = 6.7. 
Based on our reanalysis, the L/P values are now strongly clustered
between 20 and 30, with a mean of 26.6.  This is consistent with the
values which we obtained for shorter- and longer-period PRGs.

The values of $\Delta$A/\={A}, obtained by Percy and Abachi (2013),
for smaller-amplitude (1.0 down to 0.1) variables, are typically about
0.5 to 2.0.  This is consistent with the trend shown in Figure 5.
The amplitude variations are relatively larger and more conspicuous in
small-amplitude stars.

Templeton {\it et al.} (2008) call attention to three other PRGs with variable
amplitudes.  The amplitude variations in RT Hya are the largest (0.1
to 1.0) and are cyclic (L/P = 40).  The amplitude variations in W
Tau are almost as large (0.1 to 0.6) and are also cyclic (L/P = 24).
Those in Y Per are less extreme (0.3 to 0.9) and also cyclic (L/P = 29).
These three stars therefore behave similarly to PRGs in our sample.

There are therefore at least three unexplained phenomena in the pulsation of PRGs:
(1) random, cycle-to-cycle fluctuations which cause the period to
``wander"; (2) ``long secondary periods", 5-10 times the pulsation period;
and now (3) cyclic variations in pulsation amplitudes, on timescales of
20-30 pulsation periods.
PRGs have large outer convective envelopes.  Stothers and Leung (1971)
proposed that the long secondary periods represented the overturning time
of giant convective cells in the outer envelope, and Stothers (2010) 
amplified this conclusion.  Random convective cells may well explain the
random cycle-to-cycle period fluctuations, as well.  The amplitude
variations might then be due to rotational modulation, since the rotation
periods of PRGs are significantly longer than the long secondary periods
according to Olivier and Wood (2003).

\medskip

\noindent
{\bf 5. Conclusions}

\smallskip

Significant cyclic amplitude variations occurs in all of our sample of 50
mostly-Mira stars.  The relative amount of the variation (typically
Amx/Amn = 1.5) and the time scale of the variation (typically 20-35
times the pulsation period) are not significantly different in the
shorter-period and longer-period stars, and in the carbon stars.  The time
scales are consistent with those found by Percy and Abachi (2013) in a
sample of mostly smaller-amplitude SR variables, but the  {\it relative}
amplitude variations are larger in the smaller-amplitude stars.  As was
previously known: the average amplitudes increase with period for the
shorter-period stars, and the carbon stars have smaller visual amplitudes
than the oxygen stars.

\medskip

\noindent
{\bf Acknowledgements}

\smallskip

We thank the AAVSO observers who made the observations on which this
project is based, the AAVSO staff who archived them and made them publicly
available, and the developers of the VSTAR package which we used for
analysis.  This paper is based, in part, on a short summer research
project by undergraduate astronomy and physics student co-author JL.
We acknowledge and thank the University of Toronto Work-Study Program
for existing, and for financial support.  This project made use of the SIMBAD database,
maintained in Strasbourg, France.

\bigskip

\noindent
{\bf References}

\smallskip

\noindent
Benn, D. 2013, VSTAR data analysis software (http://www.aavso.org/node/803)

\smallskip

\noindent
Kafka, S. 2017, observations from the AAVSO International Database (https://www.aavso.org/aavso-international-database)

\smallskip

\noindent
Olivier, E.A., and Wood, P.R. 2003, {\it Astrophys. J.}, {\bf 584}, 1035.
 
\smallskip

\noindent
Percy, J.R., and Abachi, R. 2013, {\it J. Amer. Assoc. Var. Star Obs.}, {\bf 41}, 193.

\smallskip

\noindent
Percy, J.R., and Khatu, V.C. 2014, {\it J. Amer. Assoc. Var. Star Obs.}, {\bf 42}, 1.

\smallskip

\noindent
Percy, J.R., and Kim, R.Y.H. 2014, {\it J. Amer. Assoc. Var. Star Obs.}, {\bf 42}, 267.

\smallskip

\noindent
Price, A., and Klingenberg, G. 2005, {\it J. Amer. Assoc. Var. Star Obs.}, {\bf 34}, 23.

\smallskip

\noindent
Stothers, R.B., and Leung, K.C. 1971, {\it Astron. Astrophys.}, {\bf 10}, 290.

\smallskip

\noindent
Stothers, R.B. 2010, {\it Astrophys. J.}, {\bf 725}, 1170.

\smallskip

\noindent
Templeton, M.R. {\it et al.} 2005, {\it Astron. J.}, {\bf 130}, 776.

\smallskip

\noindent
Templeton, M.R. {\it et al.} 2008, {\it J. Amer. Assoc. Var. Star Obs.}, {\bf 36}, 1.

\smallskip

\noindent
Uttenthaler, S. {\it et al.} 2011, {\it Astron. Astrophys.} {\bf 531}, A88.

\medskip

\smallskip

\begin{table}
\caption{Pulsation Properties of Shorter-Period PRGs}
\begin{center}
\begin{tabular}{rrcrcrrcrrcl}
\hline 
Name & P(d) & MJD(1) & N & L/P & Amn & Amx & Amx/Amn & \={A} & $\Delta$A & $\Delta$A/\={A} & Note \\
\hline
T And & 281 & 16000 & 8 & 18 & 2.38 & 2.78 & 1.17 & 2.60 & 0.40 & 0.15 & s \\
V And & 256 & 20000 & 5 & 29 & 2.17 & 2.63 & 1.21 & 2.40 & 0.46 & 0.19 & s \\
UW And & 237 & 39000 & 1 & 80 & 1.68 & 2.21 & 1.32 & 2.00 & 0.53 & 0.27 & d \\
YZ And & 207 & 40000 & 3 & 28 & 2.00 & 2.51 & 1.26 & 2.25 & 0.51 & 0.23 & -- \\
S Car & 151 & 20000 & 10 & 25 & 1.03 & 1.46 & 1.42 & 1.25 & 0.43 & 0.34 & -- \\
U Cas & 277 & 20000 & 6 & 22 & 2.50 & 3.50 & 1.40 & 3.30 & 1.00 & 0.30 & s \\
SS Cas & 141 & 27500 & 5 & 43 & 1.28 & 1.78 & 1.39 & 1.55 & 0.50 & 0.32 & -- \\
Z Cet & 184 & 25000 & 7 & 25 & 2.00 & 2.45 & 1.23 & 2.25 & 0.45 & 0.20 & -- \\
T Phe & 282 & 20000 & 3 & 44 & 2.00 & 3.10 & 1.55 & 2.50 & 1.10 & 0.44 & d \\
W Psc & 188 & 40000 & 4 & 23 & 1.75 & 2.15 & 1.23 & 1.95 & 0.40 & 0.21 & s \\
RZ Sco & 160 & 25000 & 5 & 41 & 0.80 & 1.70 & 2.13 & 1.30 & 0.90 & 0.69 & d* \\
T Scl & 205 & 32000 & 2 & 63 & 1.40 & 2.40 & 1.71 & 1.70 & 1.00 & 0.59 & d \\
V Scl & 296 & 30000 & 3 & 31 & 2.05 & 2.85 & 1.39 & 2.50 & 0.80 & 0.32 & g \\
X Scl & 265 & 33000 & 6 & 15 & 1.60 & 2.03 & 1.27 & 1.80 & 0.43 & 0.24 & -- \\
S Tuc & 242 & 23000 & 6 & 24 & 2.35 & 2.85 & 1.21 & 2.65 & 0.50 & 0.39 & s \\
U Tuc & 262 & 20000 & 6 & 24 & 2.35 & 2.85 & 1.21 & 2.70 & 0.50 & 0.19 & g \\
R Vir & 149 & 20000 & 8 & 31 & 1.56 & 2.25 & 1.44 & 1.95 & 0.69 & 0.35 & -- \\
\hline
\end{tabular}
\end{center}
\end{table}

\begin{table}
\caption{Pulsation Properties of Longer-Period PRGs}
\begin{center}
\begin{tabular}{rrcrcrrcrrcl}
\hline 
Star & P(d) & MJD(1) & N & L/P & Amn & Amx & Amx/Amn & \={A} & $\Delta$A & $\Delta$A/\={A} & Note \\
\hline
R And & 410 & 20000 & 7 & 13 & 3.19 & 3.60 & 1.13 & 3.38 & 0.41 & 0.12 & g \\
X And & 343 & 16000 & 6 & 20 & 1.90 & 3.00 & 1.58 & 2.60 & 1.10 & 0.42 & d \\
RR And & 331 & 20000 & 5 & 22 & 2.68 & 3.18 & 1.19 & 3.00 & 0.50 & 0.17 & s \\
RW And & 430 & 15000 & 4 & 25 & 2.45 & 3.50 & 1.43 & 3.10 & 1.05 & 0.34 & -- \\
SV And & 313 & 15500 & 6 & 22 & 2.15 & 2.80 & 1.30 & 2.45 & 0.65 & 0.27 & -- \\
TU And & 313 & 37000 & 2 & 33 & 1.85 & 2.30 & 1.24 & 2.15 & 0.45 & 0.21 & -- \\
R Aqr & 386 & 28000 & 4 & 19 & 1.80 & 2.20 & 1.22 & 1.95 & 0.40 & 0.21 & -- \\
R Car & 310 & 20000 & 6 & 20 & 2.23 & 2.60 & 1.17 & 2.40 & 0.38 & 0.16 & -- \\
R Cas & 430 & 15000 & 5 & 20 & 2.60 & 2.98 & 1.14 & 2.73 & 0.38 & 0.14 & -- \\
T Cas & 445 & 20000 & 2 & 42 & 1.15 & 1.97 & 1.71 & 1.75 & 0.82 & 0.47 & d \\
Y Cas & 414 & 14500 & 4 & 26 & 1.80 & 2.28 & 1.27 & 2.05 & 0.48 & 0.23 & g \\
RV Cas & 332 & 20000 & 9 & 12 & 2.50 & 3.25 & 1.30 & 2.90 & 0.75 & 0.26 & s \\
TY Cas & 645 & 40000 & 1 & 27 & 2.28 & 3.20 & 1.40 & 2.90 & 0.92 & 0.32 & s \\
Y Cep & 333 & 15000 & 5 & 25 & 1.50 & 3.10 & 2.07 & 2.80 & 1.60 & 0.57 & d \\
o Cet & 332 & 20000 & 7 & 16 & 2.60 & 3.05 & 1.17 & 2.80 & 0.45 & 0.16 & -- \\
S Cet & 321 & 20000 & 5 & 23 & 2.30 & 2.87 & 1.25 & 2.70 & 0.57 & 0.21 & -- \\
W Cet & 352 & 32500 & 2 & 36 & 2.35 & 3.30 & 1.40 & 2.80 & 0.95 & 0.34 & -- \\
R Cyg & 434 & 15000 & 5 & 19 & 2.73 & 2.98 & 1.09 & 2.83 & 0.25 & 0.09 & -- \\
R Hor & 408 & 25000 & 5 & 16 & 2.95 & 3.67 & 1.24 & 3.45 & 0.72 & 0.21 & g \\
Z Peg & 320 & 20000 & 2 & 59 & 1.90 & 2.40 & 1.26 & 2.20 & 0.50 & 0.23 & d \\
\hline
\end{tabular}
\end{center}
\end{table}

\begin{table}
\caption{Pulsation Properties of Some Carbon PRGs}
\begin{center}
\begin{tabular}{rrcrcrrcrrcl}
\hline 
Star & P(d) & MJD(1) & N & L/P & Amn & Amx & Amx/Amn & \={A} & $\Delta$A & $\Delta$A/\={A} & Note \\
\hline
AZ Aur & 415 & 40000 & 2.5 & 17 & 1.35 & 1.75 & 1.30 & 1.65 & 0.40 & 0.24 & s \\
W Cas & 406 & 20000 & 3.5 & 26 & 1.18 & 1.45 & 1.23 & 1.27 & 0.27 & 0.21 & d \\
X Cas & 423 & 20000 & 5 & 17 & 0.70 & 0.93 & 1.32 & 0.80 & 0.23 & 0.28 & -- \\
RV Cen & 457 & 20000 & 3.5 & 23 & 0.83 & 1.23 & 1.48 & 1.03 & 0.40 & 0.39 & g \\
V CrB & 358 & 20000 & 6 & 17 & 1.36 & 1.75 & 1.29 & 1.50 & 0.39 & 0.26 & -- \\
U Cyg & 463 & 20000 & 4 & 20 & 1.23 & 1.55 & 1.26 & 1.45 & 0.32 & 0.22 & -- \\
T Dra & 422 & 20000 & 3 & 30 & 0.60 & 1.55 & 2.58 & 1.30 & 0.95 & 0.73 & gd \\
R For & 386 & 33000 & 5 & 12 & 1.17 & 1.53 & 1.31 & 1.35 & 0.36 & 0.27 & d \\
VX Gem & 379 & 40000 & 1.5 & 31 & 1.65 & 2.15 & 1.30 & 1.85 & 0.50 & 0.27 & gd \\
ZZ Gem & 315 & 40000 & 2.5 & 22 & 0.83 & 1.22 & 1.47 & 1.07 & 0.39 & 0.36 & g \\
R Lep & 445 & 20000 & 4 & 21 & 0.75 & 1.27 & 1.69 & 1.05 & 0.52 & 0.50 & d* \\
T Lyn & 406 & 28000 & 4 & 18 & 1.18 & 1.53 & 1.30 & 1.40 & 0.35 & 0.25 & -- \\
V Oph & 295 & 25000 & 5 & 22 & 1.03 & 1.30 & 1.27 & 1.13 & 0.28 & 0.24 & -- \\
RU Vir & 434 & 20000 & 4.5 & 19 & 1.25 & 1.78 & 1.42 & 1.40 & 0.53 & 0.38 & -- \\
R Vol & 453 & 20000 & 4 & 20 & 0.95 & 1.65 & 1.74 & 1.40 & 0.70 & 0.50 & gd \\
\hline
\end{tabular}
\end{center}
\end{table}

\begin{table}
\caption{Pulsation Properties of Some PRGs with Rapidly-Changing Periods}
\begin{center}
\begin{tabular}{rrcrcrrcrrcl}
\hline 
Star & P(d) & MJD(1) & N & L/P & Amn & A mx & Amx/Amn & \={A} & $\Delta$A & $\Delta$A/\={A} & Note \\
\hline
R Aql & 311 & 20000 & 4 & 30 & 1.83 & 2.58 & 1.41 & 2.20 & 0.75 & 0.34 & -- \\
R Cen & 502 & 20000 & 1 & 75 & 0.60 & 1.70 & 2.83 & 1.40 & 1.10 & 0.79 & d* \\
V Del & 543 & 20000 & 2 & 34 & 2.58 & 3.30 & 1.28 & 2.85 & 0.72 & 0.25 & s \\
W Dra & 291 & 20000 & 4 & 32 & 1.62 & 2.58 & 1.59 & 2.20 & 0.96 & 0.44 & -- \\
R Hya & 414 & 20000 & 3 & 30 & 1.40 & 2.25 & 1.61 & 1.70 & 0.85 & 0.50 & * \\
R Leo & 319 & 20000 & 5 & 23 & 1.60 & 2.05 & 1.28 & 1.87 & 0.45 & 0.24 & -- \\
S Scl & 367 & 20000 & 4.5 & 23 & 2.52 & 3.13 & 1.24 & 2.85 & 0.61 & 0.21 & g \\
Z Tau & 446 & 20000 & 3 & 28 & 1.45 & 2.78 & 1.92 & 1.90 & 1.33 & 0.70 & ds* \\
\hline
\end{tabular}
\end{center}
\end{table}

\begin{table}
\caption{Properties of the Amplitude Variation in Four Samples of PRGs.}
\begin{center}
\begin{tabular}{rrrrr}
\hline 
Property & SP & LP & C & CP \\
\hline
M($\Delta$A/\={A}) & 0.31 & 0.26 & 0.34 & 0.43 \\
SD($\Delta$A/\={A}) & 0.15 & 0.12 & 0.14 & 0.22 \\ 
SEM($\Delta$A/\={A}) & 0.036 & 0.028 & 0.037 & 0.076 \\ 
\hline
M(Amx/Amn) & 1.38 & 1.33 & 1.46 & 1.65 \\
SE(Amx/Amn) & 0.24 & 0.23 & 0.35 & 0.53 \\
SEM(Amx/Amn) & 0.058 & 0.052 & 0.089 & 0.188 \\
\hline
M(L/P) & 33 & 25 & 21 & 34 \\
SE(L/P) & 17 & 11 & 5 & 17 \\
SEM(L/P) & 4.1 & 2.5 & 1.3 & 6 \\
\hline
\end{tabular}
\end{center}
\end{table}

\end{document}